\documentclass[twocolumn,showpacs,preprintnumbers,prl]{revtex4}
\usepackage{graphicx}
\usepackage{dcolumn}
\usepackage{bm}
\usepackage{amsmath}
\newcommand{\kagome}{kagome }

\begin{document}
\preprint{}
\title{Frustrated minority spins in GeNi$_2$O$_4$}
\author{M.~Matsuda$^1$}
\author{J.-H.~Chung$^{2}$}
\author{S. Park$^{3}$}
\author{T. J. Sato$^{4}$}
\author{K.~Matsuno$^5$}
\author{H.~Aruga Katori$^{6,7}$}
\author{H.~Takagi$^{5,6,7}$}
\author{K.~Kakurai$^1$}
\author{K. Kamazawa$^8$}
\author{Y. Tsunoda$^9$}
\author{I. Kagomiya$^9$}
\author{C. L. Henley$^{10}$}
\author{S.-H. Lee$^{8}$}
\affiliation{$^1$Quantum Beam Science Directorate, Japan Atomic
Energy Agency, Tokai, Ibaraki 319-1195, Japan \\
$^2$Department of Physics, Korea University, Seoul, Korea\\
$^{3}$ HANARO Center, Korea Atomic Energy Research Institute, Daejeon, Korea\\
$^4$The Institute for Solid State Physics, University of Tokyo, Kashiwa, Chiba 277-8581, Japan\\
$^5$Graduate School of Frontier Sciences, University of Tokyo, Kashiwa, Chiba 277-8561, Japan\\
$^6$RIKEN (The Institute of Physical and Chemical Research), Wako, Saitama 351-0198, Japan\\
$^7$CREST, Japan Science and Technology Corporation, Saitama 332-0012, Japan\\
$^{8}$Department of Physics, University of Virginia, Charlottesville, Virginia 22904\\
$^9$Department of Applied Physics, Waseda University, 3-4-1 Okubo, Shinjuku-ku, Tokyo 169-8555, Japan \\
$^{10}$ Department of Physics, Cornell University, Ithaca, NY 14853-2501 }

\date{\today}

\begin{abstract}
Recently, two consecutive phase transitions
were observed, upon cooling, in an antiferromagnetic spinel GeNi$_2$O$_4$ at
$T_{N1}=12.1$ K and $T_{N2}=11.4$ K, respectively \cite{matsuno, crawford}. Using
unpolarized and polarized elastic neutron scattering we show that
the two transitions are due to the existence of frustrated minority spins in this compound. Upon cooling, at $T_{N1}$ the spins on the $<111>$ \kagome planes order ferromagnetically in the plane and
antiferromagnetically between the planes (phase I), leaving the spins on the $<111>$ triangular planes that separate the \kagome planes frustrated and disordered. At the lower $T_{N2}$, the triangular spins also order in the $<111>$ plane (phase II). We also present a scenario involving exchange interactions that qualitatively explains the origin of the two purely magnetic phase transitions.

\end{abstract}


\maketitle

In spinels AB$_2$O$_4$, the B sites form a highly frustrating network of corner-sharing tetrahedra, sometimes called a pyrochlore lattice.\cite{gingras,ramirez_1} In the limit of only nearest-neighbor antiferromagnetic,
isotropic exchange interactions,
this system has macroscopic classical ground state degeneracy, 
leading to a spin liquid state down to zero temperature\cite{moessner98,canals98,shlnature}, or to ordering at unobservably low temperature \cite{henley}.
The ground state degeneracy can however be lifted when the spin is coupled with lattice or orbital degrees of freedom. For instance, a spin-lattice coupling can induce a phase transition where a magnetic ordering and a lattice distortion occur simultaneously.\cite{shl2000,chung2005,hgcr2o4} When an orbital degeneracy is present, a Jahn-Teller distortion usually occurs first upon cooling to lift the orbital degeneracy. If the resulting magnetic interactions are three dimensional, then a magnetic ordering occurs simultaneously. On the other hand, if the effective magnetic interactions in the distorted phase become low dimensional, the magnetic ordering is suppressed and may occur at a lower temperature, yielding two successive phase transitions.\cite{tsunetsugu,shl2004,zhe2006} Thus, it was surprising when GeNi$_2$O$_4$ exhibited two successive phase transitions in despite of the absence of an orbital degeneracy of Ni$^{2+}$ (3$d^8$) ions.\cite{matsuno,crawford} No structural distortion was observed by synchrotron X-ray or by neutron scattering measurements, indicating that the transitions are purely magnetic \cite{crawford}.

When viewed along the $\langle$111$\rangle$ direction, the pyrochlore lattice can be described as alternating layers of a \kagome (of corner-sharing triangles) and a triangular lattice (with 1/3 as many spins per layer as the \kagome layer). We have performed polarized and unpolarized elastic neutron scattering measurements on a single crystal of GeNi$_2$O$_4$ to understand the nature of the phase transitions. Our results show that upon cooling, at $T_{N1}$ the \kagome (majority) spins order ferromagnetically in the $\langle$111$\rangle$ plane and antiferromagnetically between the planes (phase I). Our polarized elastic neutron scattering indicates that the kagome spins are aligned along a high symmetry direction in the \kagome plane. The antiferromagnetic stacking of the kagome planes induces zero internal magnetic field at the triangular spins that lie between the \kagome planes and leaves the triangular spins frustrated. The triangular spins order in the $<111>$ plane only at  the
lower $T_{N2}=11.4$ K. In order to understand the experimental findings, we have considered a spin Hamiltonian with superexchange interactions among Ni$^{2+}$ ions up to the fourth nearest neighbors. The model shows that dominant ferromagnetic $J_1$ and antiferromagnetic $J_4$ are the key interactions driving the first phase transition at $T_{N1}$, which is consistent with the Goodenough-Kanamori rules.
However, it calls for ferromagnetic $J_3$ and antiferromagnetic $J'_3$ to describe the triangular layer behavior, which does not seem to obey the rules.

A single crystal of GeNi$_2$O$_4$ was grown by chemical
vapor transport method using TeCl$_4$ as a transport agent. The
crystal has the shape of an octahedron, an edge of which is
typically about 1 mm long. Details of the sample characterization
are described elsewhere.~\cite{matsuno} The elastic
neutron-scattering experiments using unpolarized neutrons were carried out on the thermal
neutron thriple-axis spectrometer TAS2 installed at the guide hall
of JRR-3 at Japan Atomic Energy Agency. Energy of scattered neutrons was fixed to be 14.7 meV.  The horizontal collimator sequences were guide-80$^{'}$-S-80$^{'}$-80$^{'}$ or
guide-80$^{'}$-S-80$^{'}$-open. Polarized
neutron scattering experiments were carried out on the thermal
neutron thriple-axis spectrometers TAS-1 installed at
the beam hall of JRR-3. Heusler alloy (111) fixed to 14.7 meV was used as
monochromator and analyzer. Flipping ratio of
$\sim$25 was obtained on some nuclear Bragg peaks. The single crystal was mounted with the [111]
and [01\={1}] axes in the horizontal scattering plane, while a
vertical guide field was applied. Contamination from higher-order
beams was effectively eliminated using PG filters.

\begin{figure}
\includegraphics[width=9cm]{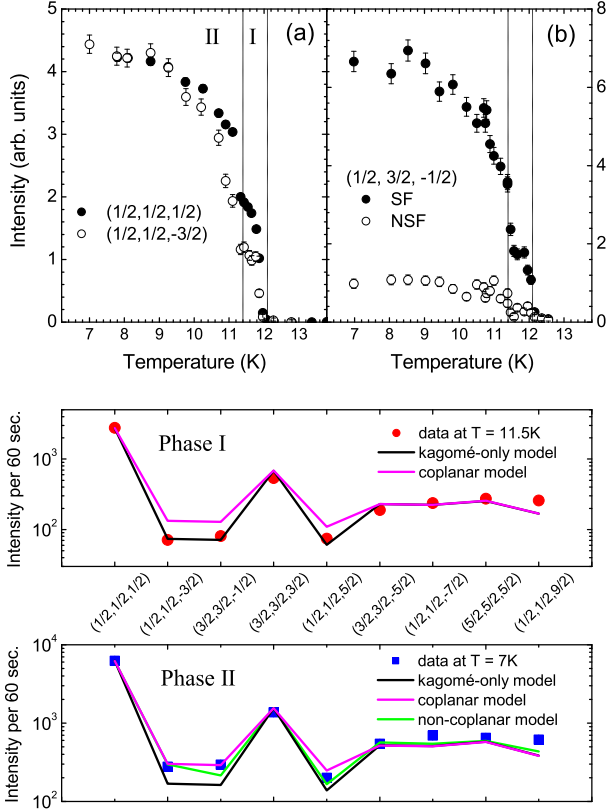}
\caption{Temperature dependence of the peak intensity of the
($\frac{1}{2}$,$\frac{1}{2}$,$\frac{1}{2}$) and ($\frac{1}{2}$,$\frac{1}{2}$,$\bar{\frac{3}{2}}$) magnetic Bragg reflections measured with unpolarized neutrons (a), and that of the ($\frac{1}{2}$,$\frac{1}{2}$,$\bar{\frac{3}{2}}$) magnetic Bragg reflections measured with polarized neutrons (b).
The vertical lines correspond to the two transition temperatures.
The experimental and calculated integrated intensities for
magnetic reflections using unpolarized neutrons at $T$=11.5 K (c) and 7 K (d). }
\label{refine}
\end{figure}

\begin{figure}
\includegraphics[width=8cm]{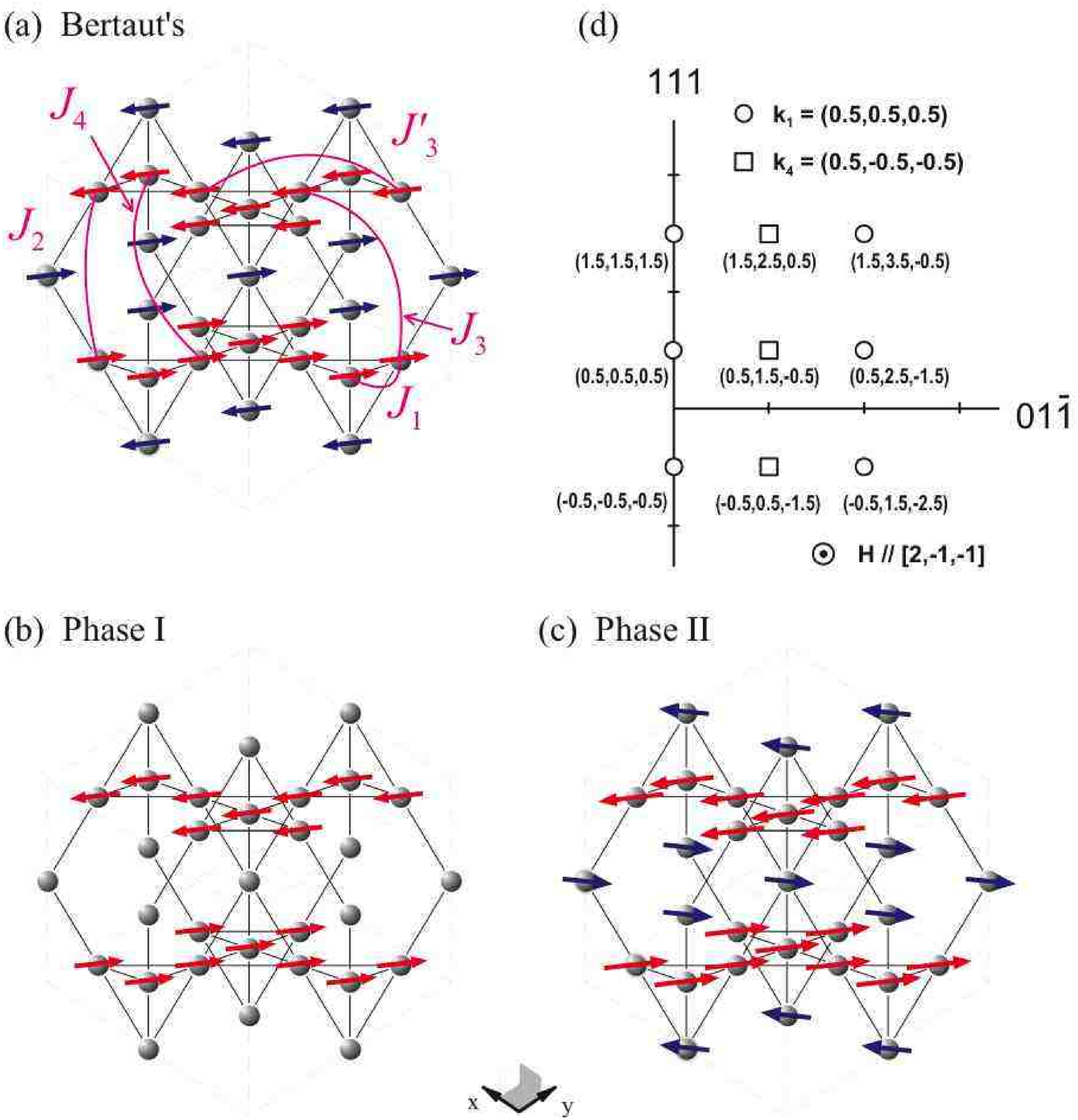}
\caption{(a) Bertaut's collinear spin model.
(b) A collinear spin model for phase I with only
\kagome spins ordered. (c) A non-collinear spin model for phase
II. All the spins are lying within the (1 1 1) plane, while the
triangular and the \kagome
spins are pointing $\sim 60^o$ away from each other. (d) The
diagram of the horizontal scattering plane for the polarized neutron measurements
measurements. The circles and squares represent the magnetic Bragg
reflections from the \textbf{k}$_1$ and \textbf{k}$_4$ domains,
respectively. The guide field was applied perpendicular to the
plane along [2\={1}\={1}].} \label{scat}
\end{figure}

Bertaut \textit{et al.} have performed neutron diffraction
measurements\cite{bertaut} on a powder sample of GeNi$_2$O$_4$ and
showed that at 4 K spins are in a long-range ordered state with a
characteristic wave vector of $\mathbf{Q}$=(1/2, 1/2, 1/2). They
have proposed as the ground state a collinear spin structure in
which, as shown in Fig. 2 (a), the spins in \kagome and triangular
planes are aligned ferromagnetically in each plane and are stacked
antiferromagnetically along the (111) direction. Our unpolarized
elastic neutron scattering on a single crystal of GeNi$_2$O$_4$
confirmed the characteristic wave vector being $\mathbf{Q}$=(1/2,
1/2, 1/2). Figure \ref{refine}(a) shows the temperature dependence
of the magnetic Bragg intensities at \it\textbf{Q}\rm$ $=
($\frac{1}{2}$ $\frac{1}{2}$ $\frac{1}{2}$) and ($\frac{1}{2}$
$\frac{1}{2}$ $\bar{\frac{3}{2}}$), obtained from the single
crystal of GeNi$_2$O$_4$. Both data show a sharp increase in
intensity around 12.1 K (phase I) with another sharp increase
around 11.4 K (phase II), which indicates the existence of two
distinct magnetic transitions and is consistent with the recent
neutron powder diffraction data \cite{crawford}. 

Our single crystal data, however,
allow us to study the nature of the phase transitions in a more
detail than the previous powder diffraction studies. The ratio of the intensities of the two reflections,
$I(\frac{1}{2},\frac{1}{2},\frac{1}{2})
/I(\frac{1}{2},\frac{1}{2},\frac{{\bar 3}}{2})$, changes as the system goes from phase I to phase II. This indicates that the phases I and II have different spin structures. In order to investigate the ordered states in detail,
we have performed unpolarized elastic scattering at nine
independent magnetic Bragg reflections at 7 K (phase II) and 11.5
K (phase I). The scans were made along the longitudinal directions
in the momentum space. Their integrated intensities are summarized
in a log scale in Fig. \ref{refine} (c) and (d). The data were
fitted to spin structures that are allowed by a linear combination
of the basis vectors of irreducible representations for the
$Fd\bar{3}m$ group with \it\textbf{Q}\rm = ($\frac{1}{2}$
$\frac{1}{2}$ $\frac{1}{2}$) \cite{izyumov, champion}. Since there
is a 3-fold symmetry within either triangular or \kagome planes, the
calculated intensities were averaged among three equivalent spin
orientations.

Interestingly, the collinear spin structure proposed by Bertaut \textit{et al} (Fig. 2 (a)) \cite{bertaut} could not reproduce our phase I data. Furthermore, none of coplanar models where all spins lie on the same plane could fit our data, as long as both \kagome and triangular spins are ordered \cite{sdomain} (see the pink line in Fig. 1 (c)).
Instead, the data was best fitted to a spin structure with only the \kagome spins ordered
with $\langle M\rangle_{kag}^I = 1.3 \mu_B$ and with the triangular spins disordered (the black line in Fig. 1 (c)).
This ordering of partial spins can be understood when we consider
the internal magnetic field at the triangular sites due to the neighboring \kagome spins.
Since the triangular plane is sandwiched by two \kagome planes below and above, the internal magnetic field due to the two
\kagome planes cancel each other on the triangular sites. Thus even when the system cools down and the \kagome spins order,
the triangular spins lying in between the \kagome planes cannot order.

When the system cools further down to phase II, the relative intensities of the magnetic Bragg reflections change (Fig. 1 (d)), indicating that the spin structure is different from that of phase I.
Since the unpolarized neutron scattering data are rather
insensitive to the orientation of the weak triangular spins, the data can be reproduced equally well by several models: any coplanar
models (see Fig. 2 (a) or (c)) with both the triangular and the
\kagome spins ordered within the $<111>$ plane with $\langle
M\rangle^{II}_{kag}=\langle M\rangle^{II}_{tri}=1.8 \mu_B$ (purple line in Fig. 1 (d)), to
which the Bertaut's model belongs, and non-coplanar models where
the triangular spins are aligned perpendicular to the $<111>$
plane whereas the \kagome spins are in the plane. In the
non-coplanar model, the best fit was obtained with the triangular
spins having smaller moments, $\langle
M\rangle_{tri}^{II}=1.3\mu_B$, than the \kagome spins, $\langle
M\rangle_{kag}^{II}=2.0\mu_B$ (green line in Fig. 1 (d)), which is more or less what a recent powder diffraction study has proposed \cite{diaz06}.
In any case, our unpolarized neutron scattering results clearly show for the first time that the two phase transitions in GeNi$_2$O$_4$ are due to changes in the triangular layers, and it makes sense that these are harder to order since their interactions are more frustrated. Our finding also explains the origin of the two distinct types of magnetic environment found in a recent muon-spin relaxation study \cite{lan06}.

Determining the actual spin directions uniquely using unpolarized neutron scattering data is difficult, if not impossible, because the intensities are usually insensitive to the orientation of the spins, as shown above. In order to distinguish the different models for the spin structure of phase II, we have performed polarized neutron measurements with a weak vertical guide field. The four possible \{111\} axes for ordering define four kinds of domains according to the angle between the ordering vector and the magnetic field: $\mathbf{k_1} = (1/2,1/2,1/2)$ with $\theta_{H,Q}$ = 90$^\circ$, $\mathbf{k_4} = (1/2,-1/2,-1/2)$ with $\theta_{H,Q}$=19.5$^\circ$, and $\mathbf{k_2} = (-1/2,-1/2,1/2)$, $\mathbf{k_3} = (-1/2,1/2,-1/2)$ with the same intermediate angle. We selected a magnetic (1/2, 3/2, $-$1/2) Bragg peak that belongs to the $\mathbf{k_4}$ domain, as the corresponding \kagome planes are closely perpendicular to the neutron polarization direction ($\sim$70.5$^\circ$) and thus the scattering in the non-spin-flip (NSF) channel is mainly due to the spin components perpendicular to the \kagome planes. (94 \% of the out-of-plane components and 34 \% of the inplane components contribute to the NSF scattering.) 
Fig. 1(b) shows the temperature dependence of the spin-flip (SF) and non-spin-flip (NSF) intensities of the reflection. The measured ratio of SF to NSF intensities is $I\rm_{exp}$(SF)/$I\rm_{exp}$(NSF)=8.7(8) after the correction for the imperfect polarization efficiency, and it stays almost constant. The measured ratio can be best reproduced when we assume all spins lie in the $\langle$111$\rangle$ plane, $I\rm_{cal}$(SF)/$I\rm_{cal}$(NSF) = 8.27, and the agreement rapidly worsens when the spins are allowed to have an out-of-plane component. So we conclude that in phase II both \kagome and triangular layer spins lie in the $\langle$111$\rangle$ plane with an unknown angle between them.


Can we interpret the ordering pattern in terms of exchange interactions? The first
five of these involve one to two Ni - O - Ni or Ni - O - Ge superexchange steps [see Fig. 2 (a) and Table I]. 
The electronic configuration Ni$^{2+} (e_g^2)$ and the Goodenough-Kanomori rules for these exchange paths imply that 
$J_1$, $J_2$, and $J'_3$ should be weak, whereas $J_3$ and $J_4$ are comparably strong and antiferromagnetic (= positive in our convention); 
Note that, as there is no structural distortion, the spin layers form through a spontaneous symmetry breaking (in which one of the four 
$\{\frac{1}{2},\frac{1}{2},\frac{1}{2}\}$ ordering vectors is arbitrarily chosen.) In fact, the ``triangular'' sites just consist of one of the four fcc sublattices of the pyrochlore structure, while the ``kagome'' sites are the union of the other three fcc sublattices. 

We considered candidate states with collinear spins in which each sublattice 
might independently have ordering vector 
$Q=\{1,0,0\}$,  $Q=\{1,1,0\}$, or $Q=\{\frac{1}{2},\frac{1}{2},\frac{1}{2}\}$ 
(either aligned or not with that sublattice's unique $\{111\}$ axis),
focusing on spin arrangements competitive with the 
observed $Q=(\frac{1}{2},\frac{1}{2},\frac{1}{2})$ pattern, 
when $J_3$, and $J_4$ are assumed strong and antiferromagnetic.
In the observed structure, the triangular and the kagome layers have 
different mean-field energies per spin:
\begin{subequations}
\begin{eqnarray}
  E_{\rm kag} & = & 2 J_1 - J_3 + J'_3 - 4 J_4;
  \label{eq:Ekag}
  \\
  E_{\rm tri} &= &3 J_3 - 3 J'_3,
  \label{eq:Etri}
\end{eqnarray}
\end{subequations}
obtained using the counts in Table~I.
This gives a net energy/spin
$E_{\rm tot} = \frac{3}{4} E_{\rm kag} + \frac{1}{4} E_{\rm tri} 
= \frac{3}{2} J_1 - 3 J_4.$
By comparison, structures built from $Q=(100)$ or $(110)$ would give
$E_{\rm tot}^* = -J_3 -J'_3 + 3 J_6 \pm | J_1 - 2 J_2 + 2 J_4|.$
Thus necessary condition to stabilize the actual structure is
$E_{\rm tot} < E_{\rm tot}^*$, implying
   \begin{equation}
       J_4  > \frac{5}{2} J_1 - 2J_2 +  J_3 + J'_3.
   \label{eq:J4J6}
   \end{equation}
which is plausible for GeNi$_2$O$_4$. 
Note that when $J_4$ dominates, the observed phase-I pattern of kagome layers 
with disordered triangular layers is the most stable mode in a mean-field approach.

As to the ordering temperatures,
mean-field  theory predicts $k_B T_{N1} = \frac{2}{3} |E_{\rm kag}|$
and $k_B T_{N2} = \frac{2}{3} |E_{\rm tri}|$, so
$T_{N2} \lesssim T_{c1}$ implies [using (\ref{eq:Ekag}) and (\ref{eq:Etri})]
that $J_4 - J_1/2 \gtrsim J'_3 - J_3$, which is consistent with the Goodenough-Kanamori rules. 
However, those rules would suggest $E_{\rm tri}\approx 3 J_3 >0$, so it is
surprising that $T_{N2}$ is close to $T_{N1}$.
Indeed, to ensure that the triangular spins order with the same
$Q=(\frac{1}{2},\frac{1}{2}, \frac{1}{2})$ as the others (rather than
the competing wavevectors), we must have $J'_3 > J_3$; $J'_3$ may be antiferromagnetic and $J_3$ ferromagnetic. 

A remaining question is, what causes {\it both} \kagome and triangular spins to point within the plane of the layers? It cannot be the uniaxial anisotropy of each spin (with respect to its local three-fold axis): 
when averaged over the three sublattices forming the \kagome spins, this must have a sign {\it opposite} to the anisotropy of the triangular spins. (since the anisotropy of all four sublattices averages to zero.)
Nor can it be interlayer Dzyalohinski-Moriya (DM) couplings: these cancel by symmetry since the spins are ferromagnetic in each layer. The simplest explanation of the observed spin orientations 
is thus dipolar anisotropy. [Anisotropic exchange couplings will also generate an effective anisotropy for either kind of spin layer, the sign of which depends on details of the anisotropy and competition amongthe exchange interactions.]

In conclusion, our analysis suggests that in GeNi$_2$O$_4$, 
$J_4$ is the key interaction driving the first phase transition at $T_{N1}$.
However, two discrepancies suggest our exchange-interaction model is
incomplete in describing the triangular layer behavior:
(i) the exchange theory would actually predict helimagnetic order with
$Q=(\frac{1}{2}+\delta, \frac{1}{2}+\delta, \frac{1}{2}+\delta)$ 
(with $|\delta| \gtrsim 0.01$), in analogy to a $J_1$-$J_2$ chain with 
$J_2 \gg |J_1|$; contrary to observation  in GeNi$_2$O$_4$ (as well as related compounds).
(ii) Exchange theory indicates FM $J_3$ and AFM $J'_3$, which is unlikely by the
Goodenough-Kanamori rules.
A complete understanding may require models with anisotropic interactions
or disorder, as well as spin-wave experiments to better constrain the $\{ J_i\}$.

\begin{acknowledgments}
We would like to thank D. Khomskii and K. Kohn for stimulating discussions
and Y. Shimojo for technical assistance. S.H.L. is supported by the U.S. DOE through DE-FG02-07ER45384, M.M. by MEXT of Japan and JSPS. and C.L.H. by NSF grant DMR-0552461.
\end{acknowledgments}

\begin{table}
\caption{\label{ratio} List of the {\it n}th nearest-neighbor interactions in a spinel, AB$_2$O$_4$, up to {\it n} = 4 and their possible {\it super}-exchange paths via anions (O) and A- or B-site transition metal ions. $J$, $d$, $\theta$, $n_B$, and $Z$ represent the coupling constant, the bond distance in a unit of $a/4$, the bond angle, the number of the exchange path, and the number of bonds, respectively. The symbol (I) stands for couplings being in $<111>$ planes, (K) and (T) for those with neighboring kagome and triangular layers, respectively. $Z_K$ and $Z_T$ are the number of bonds per a kagome  and per a triangular spin, respectively. According to Goodenough-Kanamori rules, when B ions have $e_g$ electrons, the superexchange $BOB$ is sensitive to the bond angle: it can be ferromagnetic 
(for $\theta \leq 96^\circ$) or antiferromagnetic (for $\theta \geq 96^\circ$).}
\begin{tabular}{c|c c c c | c  c}
\hline \hline
  $J$ & $d$  & Path & $\theta$  &  $n_B$& $Z_K$  & $Z_T$ \\
\hline
$J_1$ & $\sqrt{2}$ & $BOB$   & 90$^\circ$ &   2 &  4 (I), 2 (T) & 6 (K)
\\
$J_2$ & $\sqrt{6}$ & $BOAOB$ & 125$^\circ$, 125$^\circ$ & 1 & 4 (I), 4 (K), 4 (T) & 12 (K) \\
      &            & $BOBOB$ & 90$^\circ$,  90$^\circ$ & 4 &   \\
$J_3$ & $\sqrt{8}$ & $BOAOB$ & 125$^\circ$, 125$^\circ$ & 2 & 2 (I),  4 (K) & 6 (I)  \\
$J'_3$ & $\sqrt{8}$& $BOBOB$ & 90$^\circ$, 90$^\circ$ &  4  &4 (I), 2 (K) & 6 (T) \\
$J_4$ & $\sqrt{10}$& $BOAOB$ & 125$^\circ$, 125$^\circ$ & 1 & 8 (K), 4 (T) & 12 (K) \\
\hline
\end{tabular}
\end{table}


\begin{thebibliography}{}
\bibitem{sdomain} Note that the observed intensities averaged among three $S$-domains are independent of
the spin orientations within \kagome planes.
\end{thebibliography}
\end{document}